\documentclass[11pt]{article}

\usepackage[letterpaper,margin=1in]{geometry}
\usepackage{microtype}
\usepackage{graphicx}
\usepackage{subfigure}
\usepackage{booktabs}
\usepackage{tabularx}
\usepackage{amsmath}
\usepackage{amssymb}
\usepackage{algorithm}
\usepackage{algorithmic}
\usepackage{listings}
\usepackage{xurl}
\usepackage{xcolor}
\usepackage[numbers,sort&compress]{natbib}
\usepackage{hyperref}
\usepackage{footnote}
\usepackage{fancyhdr}

\hypersetup{
  colorlinks=true,
  linkcolor=blue,
  citecolor=blue,
  urlcolor=blue,
  pdftitle={Scheduling Data to Compute: Hierarchical Data Pipelining for LLM Training on AMD GPUs},
  pdfauthor={Gengyu Rao et al.}
}

\newcommand{\speedupMatrixTransposeAvg}{1.19$\times$ }
\newcommand{\speedupMatrixTransposeMax}{1.21$\times$}
\newcommand{\speedupTDMAvg}{1.88$\times$ }
\newcommand{\speedupTDMMax}{2.74$\times$}

\newcommand{\speedupSDMAAvg}{1.16$\times$ }

\newcommand{\speedupRDMAAvg}{1.33$\times$ }

\title{\textbf{Scheduling Data to Compute: Hierarchical Data Pipelining for LLM Training on AMD GPUs}}

\author{\small
Gengyu Rao, Ghadeer Ahmed H Alabandi, Su Ann Chong, Thomas Huber,\\
Kailash Gogineni, Subhajit Dutta Chowdhury, Saleel Kudchadker, Suchita Pati,\\
Zirui Zhang, Zhen Huang, Yanyuan Qin, Xiaoming Peng, Lihuan Zhang, Botao Hu,\\
Wei Huang, Yang Bo, Wen Xie, Xiangyu Zhao, Ruibin Zhang, Xiaobo Chen,\\
Vidushi Goyal, Anshu Raina, Claire Songhyun Lee, Jiakun Yan, Yao Fu,\\
Nusrat Islam, Ravi Dwivedula, Jie Zhang, Maria Garzaran, Zhenyu Gu, Emad Barsoum\\[0.6em]
\normalsize Advanced Micro Devices, Inc.\\
\texttt{lorrirao@amd.com}
}
\date{}

\begin{document}
\maketitle

\begin{abstract}
Large language models (LLM) have been the key focus of recent machine learning breakthroughs. As the model size grows exponentially, the computation needed for training is astronomical. A GPU can do a huge amount of arithmetic. However, the data its compute units need is not always available when they are ready. This data may still be in slow memory or on another GPU. While the chip waits, expensive compute sits idle, and at scale even a small drop in utilization translates into substantial cost. In this paper, we treat this waiting time as the primary bottleneck. We focus on hardware features designed for explicit data movement. These features load data from off-chip memory onto the chip, rearrange it into the layout the computation requires as it is loaded, and transfer it between GPUs. By orchestrating them, we keep a steady pipeline of data flowing to the compute units so that they remain busy.

We present kernel and framework implementations that address this bottleneck using three key data-movement instructions and hardware features. The Tensor Data Mover (TDM) moves tensor data from high-bandwidth memory (HBM) into on-chip memory. Transpose-load instructions transpose tensor data into the layout that LLM computations require as it is loaded on-chip. Finally, System DMA (SDMA) and Remote DMA (RDMA) move data between GPUs over the scale-up and scale-out networks, respectively.
Our results show that TDM and transpose-load provide on average \speedupTDMAvg and \speedupMatrixTransposeAvg speedup at the kernel level, respectively. The SDMA and RDMA lead to on average \speedupSDMAAvg and \speedupRDMAAvg speedup for computation overlapped with communication. This is the first AMD LLM training system to utilize all the GPU data pipelining features. Our implementation has been integrated into open source framework and library for AMD GPUs. 
{\url{https://github.com/AMD-AGI/Primus-Turbo}} {\url{https://github.com/AMD-AGI/Primus}}

\end{abstract}

\section{Introduction}
\label{intro}
Recent years have witnessed an exponential surge in LLMs. People are training larger and larger models with an unprecedented amount of data. As the training grows exponentially, the amount of computation needed is unprecedented in human history. People are building larger and larger GPU clusters to train LLM models, due to both the model size and the amount of computation. There have been many reports on GPU under-utilization in the industry \cite{lu2026bridging,makin2025sustainable,meta-scale}. For hyper-scale users, even one percent of efficiency can mean millions of dollars. From a hardware perspective, under-utilization can arise from several sources, including hardware failures \cite{kokolis2025revisiting}, power constraints \cite{zhao2023sustainable}, and data-movement overheads \cite{ivanov2021data,wang2024zero}, which become increasingly significant as models and distributed training systems scale.

In this paper, we focus on data-movement overhead as a source of GPU under-utilization. Within a GPU, compute units may stall while waiting for data or spend execution resources moving and transforming it. Across GPUs, communication may consume compute resources or delay dependent computation. We consider these two cases separately as intra-GPU and inter-GPU data movement.


\noindent\textbf{Intra-GPU data movement.} Traditional GPU programming emphasizes that GPUs hide long memory-access latency through inexpensive thread switching, i.e., scheduling more threads than the hardware can execute at once \cite{volkov2016understanding}. In recent years, however, \textbf{software-managed data pipelines} have become an increasingly important complementary approach to latency hiding.

A key driving force behind this shift is the co-evolution of GPU hardware architectures and kernel algorithm design. Consider attention kernels, underpinning the Transformer architectures central to recent breakthroughs in large language models (LLMs) \cite{vaswani2017attention}. Modern high-performance attention kernels largely follow the approach introduced by FlashAttention \cite{flashatten}: keeping the softmax state and output accumulators on-chip while streaming key and value tiles from memory. Although this approach avoids materializing the full attention matrix and significantly reduces global memory traffic, it introduces a new performance challenge of efficiently orchestrating data movement between global and on-chip memory. In particular, attention computation requires fine-grained scheduling of KV tile loading, data transformation, and computation to prevent memory latency from stalling the execution units.

On older GPU architectures, transferring data into on-chip memory requires explicit load instructions executed by threads on the compute units. As a result, data movement competes with computation for execution resources. While one group of threads waits for memory operations to complete, another group performs useful computation.  However, this approach becomes increasingly challenging for attention kernels, where limited on-chip memory capacity and frequent data transfers demand tighter coordination between memory operations and computation.

Recent GPU architectures are beginning to address this challenge. For example, AMD's MI455 introduces the Tensor Data Mover (TDM), which moves data between HBM and on-chip memory. This not only allows software-managed data pipelines to be offloaded to hardware DMA engine, but also introduces new fine grain instruction set architecture (ISA) to coordinate between memory transfer and computation, fundamentally changing the optimization target from “launch enough threads to hide a high-latency load” to “build a steady stream that keeps the compute units fed."

A similar trend can be observed in data transformation. During LLM training, tensors frequently require different layouts for forward and backward computations, making matrix transposition an important source of overhead. On MI355, specialized matrix-transpose instructions allow data to be transposed during loading, eliminating the need for explicit register-level transposition by the compute units \cite{amd-cdna4-isa}.

We will detail our implementation and the performance impact of these ISA changes in Section \ref{sec:intra-gpu-data-movement}.
Our implementation shows that matrix transpose leads to on average \speedupMatrixTransposeAvg and up to \speedupMatrixTransposeMax. TDM leads to on average \speedupTDMAvg and up to \speedupTDMMax. 


\noindent\textbf{Inter-GPU data movement.} As large-scale LLM training expands to thousands of GPUs, inter-GPU data movement has become an increasingly important performance bottleneck. At first glance, overlapping communication with computation appears straightforward: ideally, network transfers proceed concurrently with GPU computation, achieving both high network bandwidth and high compute utilization. In practice, however, communication and computation often compete for the same GPU execution resources, making effective overlap much more challenging.

For example, traditional RCCL-based communication from AMD can occupy up to 64 compute units (CUs), corresponding to 25\% of the compute resources on MI355. Consequently, even when communication is fully overlapped with computation, we can observe significant slowdown in compute throughput. This reveals a fundamental limitation of conventional GPU-driven communication: hiding communication latency does not necessarily eliminate its computational overhead.

One possible workaround is to increase the local batch size, thereby performing more computation per communication operation and amortizing communication overhead. However, this approach conflicts with the growing demand for faster distributed training. For customers renting GPU clusters in the cloud, time-to-train is often as important as overall hardware efficiency. For example, assuming near-linear scaling, training a model on more than 2,000 GPUs in two months may be preferable to training it on 1,000 GPUs over four months. Achieving such scaling requires maintaining high efficiency at smaller per-GPU batch sizes, where communication accounts for a larger fraction of execution time. Therefore, simply increasing the amount of local computation is not always a viable solution.


To address this challenge, AMD GPUs and Network Interface Cards (NICs) provide dedicated DMA engines for inter-GPU data movement, including System DMA (SDMA) and Remote DMA (RDMA). Unlike traditional communication kernels that consume GPU compute resources, these engines enable data transfers to proceed independently of the compute units, creating opportunities for more effective communication–computation overlap. To make these capabilities accessible, we have been introducing lightweight, DMA-based communication APIs across multiple layers of the ROCm software stack. These APIs expose dedicated data-movement engines through programmer-friendly interfaces, supporting both high-level collective operations in PyTorch and low-level, customized communication implementations.


We will provide detailed evaluation of overlapping overhead in Section~\ref{sec:sdma_eval} and detail our hardware support and software API in Section~\ref{sec:impl}. 
Our results show that System and Remote DMA based optimization leads to on average \speedupSDMAAvg and \speedupRDMAAvg of overlapped computation with communication.



\section{Background and motivation} 
\subsection{Attention Kernel}

Attention is a core operation in modern large language models \cite{vaswani2017attention}. For a sequence of
$T$ tokens, each attention head projects the input activations into the query, key, and value matrices
$Q,K \in \mathbb{R}^{T \times d}$ and $V \in \mathbb{R}^{T \times d_v}$, where $d$ and $d_v$
denote the key/query and value dimensions, respectively. Scaled dot-product attention computes:
\begin{equation}
  O = \operatorname{softmax}\left(\frac{QK^{T}}{\sqrt{d}}\right)V,
\end{equation}

Most current implementations follow FlashAttention\cite{flashatten}, as illustrated in Algorithm~\ref{alg:flash-attention}:
It streams blocks of $K$ and $V$ through on-chip memory, updates the online softmax state, and output accumulator.
This algorithm makes the timing of the streamed $K,V$ tiles critical and turns attention optimization into a tiled dataflow problem: we need to fetch each tile of $K$ and $V$ in time, so that the computing units are always busy.


\setcounter{algorithm}{-1}
\begin{table}
\centering

\small
\refstepcounter{algorithm}
\noindent\rule{\linewidth}{0.4pt}
\textbf{Algorithm \thealgorithm: FlashAttention-style attention \cite{flashatten}.}
\label{alg:flash-attention}
\begin{algorithmic}[1]
\REQUIRE $Q,K\in\mathbb{R}^{T\times d}$, $V\in\mathbb{R}^{T\times d_v}$ in HBM;
query and key/value block sizes $B_q,B_k$
\STATE Partition $Q$, $K$, and $V$ into blocks.
\STATE Initialize $O_i\gets0$, $m_i\gets-\infty$, and $\ell_i\gets0$.
\FOR{each key/value block $(K_j,V_j)$}
\STATE Load $K_j,V_j$ from HBM into on-chip memory.
\FOR{each query block $Q_i$}
\STATE Load $Q_i,O_i,m_i,\ell_i$ into on-chip memory.
\STATE On chip, use $Q_iK_j^T/\sqrt{d}$ to update $O_i$, $m_i$, and $\ell_i$
       with online softmax normalization and $V_j$.
\STATE Write the updated $O_i$, $m_i$, and $\ell_i$ to HBM.
\ENDFOR
\ENDFOR
\STATE \textbf{return} $O$
\end{algorithmic}
\noindent\rule{\linewidth}{0.4pt}
\end{table}

To make this issue more complex, as the industry is starting to use longer and longer sequence lengths, more attention variants have been proposed to address the HBM memory bandwidth issue:

\begin{itemize}
\item \textbf{Sparse Attention}: It restricts each query to a selected subset of keys and values, and can also reduce memory traffic and computation. 
\item \textbf{Compressed Attention}: It compresses token information, thereby reducing both memory traffic and computation.
\end{itemize}

The recent DeepSeek V4 model proposed Compressed Sparse Attention (CSA), which uses the two optimizations mentioned above \cite{xu2026deepseek}. 
This creates additional challenges to kernel implementation, as the dataflow is more irregular: Each query
token receives a top-$k$ list containing a sliding-window component and, depending on the compression
ratio, additional compressed latent vectors. The forward pass must gather those vectors, compute scores,
and stream the corresponding values into the output accumulator. The backward pass is more demanding:
it must gather the same sparse vectors to form $dQ$, produce a per-token/per-slot intermediate for
$dKV$, and then reduce that intermediate by the inverse of the top-$k$ map \cite{liu2025deepseek,xu2026deepseek}. 

The kernel case study in this paper focuses on this new sparse attention algorithm, as it represents one of the most challenging data movement patterns in LLM training and represents a possible direction for the future evolution of attention algorithms.

\subsection{Communication for LLM training}
In recent years, as LLM sizes and architectures evolve, parallelism and communication pattern have also changed. The traditional parallelism strategy includes data parallelism and data sharding, including ZeRO sharding, FSDP \cite{zero, pytorch-fsdp}. Correspondingly, the communication patterns are AllReduce, AllGather and ReduceScatter. 
Other forms of parallelism include tensor parallelism (TP) \cite{shoeybi2019megatron}, context parallelism (CP) \cite{niu2026efficientscalingllmtraining}, etc. As the MoE architecture has become more popular, targeting parallelism and communication patterns have also emerged (e.g., DeepEP \cite{deepep2025} and the All2All communication), also more recently MegaMoe kernel \cite{megamoe}, which fused communication with computation. In this paper, we focus on the traditional FSDP pattern. MoE patterns have been studied by previous AMD work \cite{rocm-deepep, primus-turbo-megamoe}.

\section{AMD GPU architecture features and workload implementations}
\label{sec:impl}
In this section, we will outline AMD GPU architecture features, and how we utilized them in our kernel and framework implementation. Since this paper is focused on case studying GPU architecture from a workload perspective, we will focus on the ISA and API exposed to the user, instead of in-depth discussion of the hardware design itself. 
\subsection{Intra-GPU data movement}
\label{sec:intra-gpu-data-movement}

\begin{figure}
    \centering
    \includegraphics[width=0.45\textwidth]{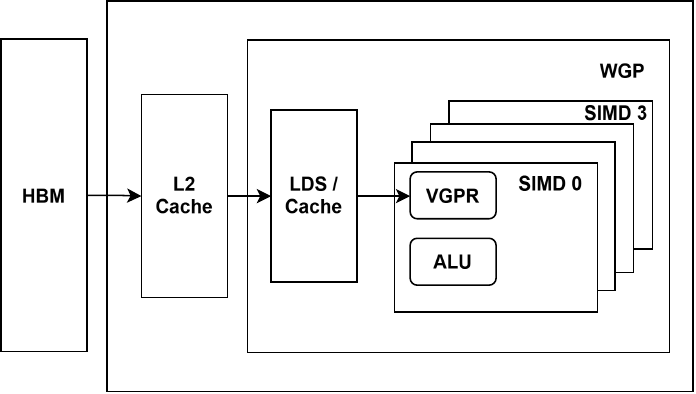}
    \vspace{-4mm}
    \caption{Simplified memory hierarchy of AMD CDNA5 GPU \cite{amd-cdna5-isa}.}
    \vspace{-2mm}
    \label{fig:gpu-memory-arch}
\end{figure}

The AMD GPU architecture is illustrated in Figure~\ref{fig:gpu-memory-arch}. 
Table~\ref{tab:amd-nvidia-terminology} compares the ROCm terms used in this
paper with the closest CUDA terminology. The correspondences explain
programming roles, rather than claiming that the underlying hardware
implementations are identical. Since CUDA programming is more popular in GPU research literature, we will use these terms interchangeably in this paper.

For our memory hierarchy, the \emph{Local Data Share} (LDS) is a low-latency on-chip memory shared by the threads (Work-item) in a workgroup. A kernel uses LDS to stage and reuse a tile that would otherwise be read repeatedly from HBM; threads (Work-items) coordinate accesses with workgroup barriers. LDS is
allocated per workgroup, so its contents are not visible to other workgroups.

\begin{table*}[t]
\centering
\caption{ROCm and CUDA GPU programming terminology \cite{amd-cdna5-isa,cuda}.}
\label{tab:amd-nvidia-terminology}
\begin{tabular}{p{0.18\textwidth}p{0.19\textwidth}p{0.55\textwidth}}
\toprule
ROCm term & Similar CUDA term & Meaning \\
\midrule
\midrule
Work-item & Thread & One logical execution flow of a kernel. \\
\midrule
Wave & Warp & A collection of 32 work-items/threads that execute in parallel. MI455 (CDNA5) supports 32 threads; the earlier MI355 (CDNA4) supports 64 threads\\
\midrule
Workgroup & Thread block & A collection of waves/warp that can synchronize with each other using barriers; \\
\midrule
Local Data Share (LDS) & Shared memory & A multi-bank on-chip scratch memory allocated to waves or work-groups. \\
\midrule
Vector General Purpose Registers (VGPR) & Register & Per-work-item register storage, allocated from hardware register resources. \\
\midrule
Tensor Data Mover (TDM) & Tensor Memory Accelerator (TMA) & DMA transfers of matrix/tensor data. \\
\midrule
SIMD/Compute Unit (CU)/Work-group Processor (WGP)  & Streaming Multiprocessor (SM) & A hardware unit that executes workgroups. In MI455, each WGP consists of two CU and four SIMD. Every SIMD can execute a wave of 32 threads. \\

\bottomrule
\end{tabular}
\end{table*}

Specialized ISA and hardware have been introduced on AMD GPU to handle data-movement operations.
These include matrix transposing ISA and Tensor Data Mover (TDM), both would free the useful compute cycle from data-movement operations. 
\subsubsection{Tensor Data Mover}
\label{sec:tdm_impl}
Table~\ref{tab:tdm-programming-model} shows the ISA/intrinsic provided by MI455 for moving data between HBM and on-chip memory. \texttt{TENSOR\_LOAD\_TO\_LDS} and \texttt{TENSOR\_STORE\_FROM\_LDS} are used to load and store data from HBM to on-chip memory (LDS) and from LDS to HBM, respectively. These DMA engines are aware of the tensor data structure: they are able to fetch data based on tensor dimensions and can also pad the data to resolve LDS bank conflicts.

\texttt{TENSORCNT} counts the number of matrix (tensor) transfers outstanding but not yet completed. \texttt{S\_WAIT\_TENSORCNT} waits until the number of outstanding TDM transfers is not greater than
the specified threshold. For a pipeline of depth N, we can start the pipeline with N+1 tensor loads and simply check \texttt{S\_WAIT\_TENSORCNT} with N. Because tensor-movement instructions complete in-order with other Tensor instructions from the same wave (both loads and stores stay in-order), we can ensure the data is ready without worrying about memory transfer ordering \cite{amd-cdna5-isa}. 

The important programming property is that the gather is not expressed as a loop of vector global loads into LDS. And the programming mindset changes from ``launch enough workgroups to hide a high-latency load'' to ``build a steady stream that keeps the compute pipeline fed.'' This also lowers the occupancy and frees more registers for the kernels, so more data can be reused in registers.

In addition, \texttt{TENSOR\_LOAD\_TO\_LDS} also supports sparse gather. One load instruction takes up to eight 32-bit or sixteen 16-bit indices. The TDM engine fetches each indexed memory range from the HBM and writes them to LDS. Padding can also be added via this instruction to resolve bank conflicts.

This matches the sparse attention algorithm directly, because the top-$k$ list of a token is exactly a list of KV row indices. Our kernels read eight indices with one scalar load and pass them to \texttt{TENSOR\_LOAD\_TO\_LDS}. Without gather mode, the same load would require a loop of per-lane address computation and explicit load instruction from HBM to LDS. This would waste both registers and execution resources that the computation needs.

We used these ISA to construct our Compressed Sparse Attention (CSA) kernels. The CSA attention has three variants: \texttt{cr0}, \texttt{cr4}, and \texttt{cr128} stand for compression ratios of 0, 4 and 128. Please refer to Table \ref{tab:tdm-benchmark-matrix} for further explanation.

\begin{table*}[t]
\centering
\caption{TDM ISA/intrinsic \cite{amd-cdna5-isa}.}
\label{tab:tdm-programming-model}
\begin{tabularx}{\textwidth}{@{}p{0.34\textwidth}X@{}}
\toprule
ISA/intrinsic & Purpose \\
\midrule
\texttt{TENSOR\_LOAD\_TO\_LDS} & Load data from HBM to on-chip memory \\
\texttt{TENSOR\_STORE\_FROM\_LDS} & Store data from on-chip memory to HBM\\
\texttt{TENSORCNT} & Counts the number of matrix (tensor) transfers outstanding
but not yet completed \\
\texttt{S\_WAIT\_TENSORCNT} & Waits until number of unfinished transfers is less-than or equal-to specified value\\
\bottomrule
\end{tabularx}
\end{table*}

The forward kernel uses TDM to stream tiles from the top-$k$ key-value entries: it uses  \texttt{TENSOR\_LOAD\_TO\_LDS} with a vector of addresses to KV tiles and an LDS destination, then syncs via \texttt{S\_WAIT\_TENSORCNT} before consuming the completed tile.  
For each token, the kernel first
issues TDM load for the KV tiles, then computes the $QK$ and $PV$ on the earlier tile,
meanwhile issues TDM load for a later tile.

For the \texttt{cr0} / \texttt{cr128} sliding-window kernels, the tiles are reused across tokens: consecutive tokens attend to almost the same window rows, so a workgroup owns a block of consecutive tokens, gathers their shared window (and, for \texttt{cr128}, the compressed pool) into LDS once with TDM, and each token reads its 128 window rows at its own offset instead of gathering them again.

For \texttt{cr0} / \texttt{cr128} with short top-$k$, our kernel implementation makes the workgroup \emph{persistent}
across tokens: the global tile index continues across token boundaries. This removes part of the per-token prologue from the critical path. For \texttt{cr4} with a longer top-$k$, the prologue is already amortized, so our kernel implementation with a straight-forward token-fixed kernel already achieves good hardware utilization. The persistent kernel leads to additional overhead.

The backward kernel uses TDM in the same way, with some backward-specific changes. We will evaluate TDM's impact on kernel performance in Section~\ref{sec:tdm-eval}.
\subsubsection{Matrix Transposing}
\label{sec:matrix-trans-impl}

For transposing, Table~\ref{tab:transpose-load-interface} shows the ISA interface used by our kernels. For MI455, all the 32 threads in a wave/warp will execute the \texttt{ds\_load\_tr16\_b128}. In total, they will load and transpose a $16\times16$ matrix of 16-bit data. We directly use the transpose-load ISA when feeding tensor fragments from on-chip memory to registers. 

Without this transpose-load ISA, we have to use XOR-based register shuffle to match the Matrix Multiply Accumulate operand layout, which wastes both register resources and compute cycles. Or we will have to use strided-load from on-chip memory into register, which will waste bandwidth from on-chip memory to registers.
Table~\ref{tab:transpose-load-demo} shows the kernel implementation with and without transpose load support. 

This data transposing is a challenge for data movement in LLM training and is less a problem in LLM inference. 
As in inference, we can pre-transpose weight tensors offline for direct computation at runtime. However, training performs both forward and backward passes and uses different layouts. 
To explain the training process, we list part of the attention tensor layout in Table \ref{tab:attention-transpose-shapes}. For example, even if we may transpose the $K$ data in forward pass for direct computation, in the backward pass, we will still have to transpose it back.

\begin{table*}[t]
\centering
\caption{Attention kernel computes and layout in training}
\label{tab:attention-transpose-shapes}
\begin{tabular}{llll}
\toprule
Pass & Operation & Tile shapes & Transposed operand \\
\midrule
Forward & Scores & $Q_i[B_q,D_q] \times K_j^T[D_q,B_k]$ & $K_j^T$ \\
Backward & Query gradient & $dS_{ij}[B_q,B_k] \times K_j[B_k,D_q]$ & none \\
Backward & Key gradient & $dS_{ij}^T[B_k,B_q] \times Q_i[B_q,D_q]$ & $dS_{ij}^T$ \\
\bottomrule
\end{tabular}
\end{table*}

We will also evaluate its performance impact in Section \ref{sec:transpose-eval}. Even though we only discussed the attention kernel, the same ISA also applies to kernels like GEMM.

\begin{table*}[t]
\centering
\caption{Transpose-load ISA / intrinsic \cite{amd-cdna5-isa,amd-cdna4-isa}.}
\label{tab:transpose-load-interface}
\small
\begin{tabularx}{\textwidth}{@{}l p{0.25\textwidth} p{0.15\textwidth} X@{}}
\toprule
GPU & ISA / intrinsic & Transpose load size & Description \\
\midrule
MI355X & \texttt{ds\_read\_b64\_tr\_b16} & 64-bits per thread & Used for either column major matrix A or row major matrix B data load to 2 VGPRs. Element size is 16b. Two instruction load a complete matrix \\\midrule
MI455X & \texttt{ds\_load\_tr16\_b128} & 128-bits per thread & Load a $16\times16$ matrix of 16-bit data into VGPRs and transpose between row-major and column-major order. This instruction loads data into 4 consecutive VGPRs. A wave/warp of 32 threads moves $32 \times 128 $ bits.\\
\bottomrule
\end{tabularx}
\end{table*}

\begin{table*}[t]
\centering
\caption{matrix transpose with / without LDS transpose
load.}
\label{tab:transpose-load-demo}
\begin{tabular}{p{0.47\textwidth}p{0.47\textwidth}}
\toprule
Manual transpose in registers & Transpose-load ISA \\
\midrule
\begin{minipage}[t]{0.47\textwidth}
\small\ttfamily
rows = lds\_load(tile\_ptr)\\
tmp0 = xor\_permute(rows, 1)\\
tmp1 = xor\_permute(tmp0, 2)\\
tmp2 = xor\_permute(tmp1, 4)\\
b\_frag = pack\_wmma\_operand(tmp2)\\
acc = wmma(a\_frag, b\_frag, acc)
\end{minipage}
&
\begin{minipage}[t]{0.47\textwidth}
\small\ttfamily
b\_frag = lds\_load\_tr16(tile\_ptr)\\
acc = wmma(a\_frag, b\_frag, acc)
\end{minipage}
\\
\bottomrule
\end{tabular}
\end{table*}

\subsection{Inter-GPU communication}
\label{sec:overlap_slow_down}

In a typical GPU cluster, there are two types of GPU networks: scale-up and scale-out networks. 
They differ in terms of physical distance and communication bandwidth/latency.

Scale-up networks are often used to connect GPUs within a server rack/blade, which are physically close and enjoy better network bandwidth (per GPU) and latency.

Scale-out networks are used to connect GPUs across server racks or blades. Due to the signal integrity and power issues caused by physical distance, the scale-up network technology usually does not apply \cite{meta-network}. In a typical GPU cluster, the scale-out bandwidth per GPU is much smaller than the scale-up bandwidth, and the latency is also typically higher.

In this section, we will first discuss inter-GPU communication within a scale-up domain, then we will extend to scale-out communication.

\subsection{Scale-up domain: System DMA (SDMA)}
\begin{figure}[t]
\centering
\includegraphics[width=\columnwidth]{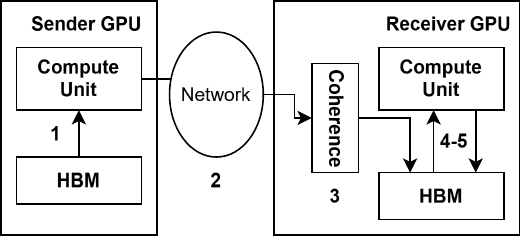}
    \vspace{-2mm}
\caption{RCCL communication collective memory access illustration.}
    \vspace{-2mm}
\label{fig:ll-protocol-hbm}
\end{figure}

In traditional communication implementation within a scale-up domain, as shown in Figure~\ref{fig:ll-protocol-hbm}, a communication would begin with GPU threads reading from local HBM (as shown in Step 1), and writing through the scale-up fabric (as in Step 2). The remote GPU's Coherence Station (CS) will write into a staging buffer in the HBM (as in Step 3). Then the remote GPU threads will be used to read the data from staging buffer and write into the destination buffer (as in Step 4 and 5).   
This design would waste a significant amount of GPU threads / compute units, preventing them from doing useful computation. In MI355, by default, the wasted Compute Units will be 25\% of the total Compute Units (64 out of 256).

From a power perspective, this implementation triggers unnecessary data movement across chiplets, which leads to increased power usage. AMD GPUs provide dedicated on-chip units called SDMA engines to handle those data movement.
Previous AMD research \cite{pal2025design} showed that this data movement offloading can reduce GPU power usage. This is very important because, during LLM training, we found that the compute and communication may compete for power: When the DMA transfer began, we observed drops in the compute-die frequency and power.

These DMA engines can be initiated by either GPU or CPU:
For CPU programming, those SDMA engines are exposed to user through the HIP APIs \texttt{hipMemcpyBatchAsync} and \texttt{hipMemcpyAsync}. For GPU programming, they are ready to use via ROCm libraries: RCCL (GIN mode), MORI and rocSHMEM \cite{rccl,mori,rocshmem}.

To use this DMA feature in LLM training, before initiating those DMA transfers, we will first need to let the DMA engine know where the remote memory physically locates. To do this, we will first need to allocate a local memory and obtain its handle, then we will need to transfer the handle to remote GPU and let the remote GPU maps the address to its virtual address. 

For user simplicity, all those processes are hidden under different ROCm and PyTorch stacks. For PyTorch users, people will only need to register the memory buffer via the symmetric memory API, which would handle the underlying memory mapping. Then the DMA based communication can be triggered for data write/read on those mapped remote memory. 

For users who want to define their own parallelism and communication pattern, the whole process can be easily implemented via tens of lines of PyTorch code, as illustrated in Code~\ref{lst:pytorch-sdma}. For users who want to directly use the FSDP parallelism, this process has been fully integrated into the PyTorch framework and can be easily enabled by a few lines, as illustrated in Pseudo Code~\ref{lst:pytorch-fsdp}. All the latest PyTorch release builds for AMD GPU should have this function ready to use.

\begin{lstlisting}[
  language=Python,
  float=tp,
  caption={Pseudo PyTorch code for customized communication using SDMA.},
  label={lst:pytorch-sdma}
]
import torch.distributed as dist
import torch.distributed._symmetric_memory as symm_mem
options = dist.ProcessGroupNCCL.Options()
cta_policy = 2 # zero-CTA
_SDMA_GROUP = dist.new_group(
            ranks=list(range(dist.get_world_size())),
            backend="nccl",
            pg_options=options,
            group_desc=f"SDMA demo",
        )
symm_mem.set_backend("NCCL")
symm_mem.enable_symm_mem_for_group(_SDMA_GROUP)
dist.barrier(group=group)
pool = symm_mem.get_mem_pool(device)
with torch.cuda.use_mem_pool(pool):
    storage = torch.empty(size_bytes, dtype=torch.uint8, device=device)
torch.distributed.all_gather_into_tensor(
                        output,
                        local_input,
                        group=group,
                        async_op=True,
                    )
\end{lstlisting}

\begin{lstlisting}[
  language=Python,
  caption={PyTorch FSDP using SDMA.},
  label={lst:pytorch-fsdp}
]
from torch.distributed.fsdp._fully_shard._fsdp_collectives import SymmMemAllGather
fsdp_module.set_custom_all_gather(SymmMemAllGather(pg, backend))
\end{lstlisting}

We've integrated these new DMA based communication into our LLM training framework. Users can enable the DMA based communication for LLM training via simple ENV flags. We will evaluate them in section \ref{sec:eval}.

\subsection{Scale-out domain: Remote DMA (RDMA)}
Different from the scale-up domain, in the scale-out domain, the GPU cannot directly access each other's memory and must rely on the Network Interface Card (NIC) to perform the data transfer operation.
The DMA transfer in the scale-out domain is called RDMA. To enable GPU initiated communication, AMD has implemented GPU Initiated Networking (GIN) via CPU proxy threads, which will work as a delegate for the GPU to initiate the RDMA transfer.

\begin{figure}[t]
\centering
\includegraphics[width=0.7\columnwidth]{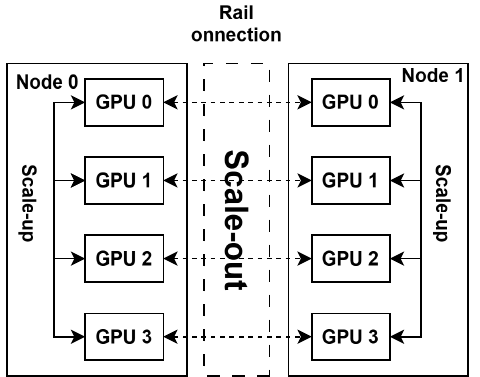}
\caption{Example GPU cluster networking topology.}
\label{fig:rdma-topo}
\end{figure}

Recent AMD works have enabled chaining SDMA and RDMA to perform communication across scale-up and scale-out domains. Figure \ref{fig:rdma-topo} shows a typical GPU cluster networking topology. In GPU networking, GPUs on the same index across different nodes are called as on the same "rail". The hierarchical DMA transfer can be explained as below: All the GPUs on the same "rail" will first perform RDMA transfer (as dotted lines in the figure). And then, each GPU will use local SDMA to broadcast the data it received (as the solid line in the figure). For large data transfer, this RDMA and SDMA will be overlapped, as each GPU sends chunked data as pipeline via RDMA and SDMA at the same time. Thus, all the scale-up and scale-out network bandwidth can be utilized.

The complexity has been hidden from users within the ROCm stack, and users can directly utilize the feature via the same PyTorch interface mentioned in the Scale-up section.

\section{Evaluation}
\label{eval}
\label{sec:eval}

\subsection{Hardware platform}
We implement our performance optimizations in Primus~\cite{primus} and Primus-Turbo~\cite{primus_turbo} to maximize GPU utilization, enhance computational efficiency, and accelerate large-scale AI model training on AMD GPUs. We evaluate these optimizations on two AMD GPU architectures, MI455 and MI355, whose hardware specifications are compared in Table~\ref{tab:455vs355}.

For Intra-GPU ISA, we evaluate on MI455, as it supports all the data moving ISA we mentioned above. All the MI455 results below are collected from bring-up platform. As the MI455 platform continues to undergo performance optimizations, the reported results reflect the best implementations available at the time of evaluation and should be interpreted as a point-in-time snapshot rather than definitive performance limits. For Inter-GPU communication, we will evaluate on MI355, which is currently widely used in industry.

\begin{table}[t]
\centering
\caption{MI455 vs MI355 architecture feature comparison \cite{amd-cdna5-isa,amd-cdna4-isa}.}
\label{tab:455vs355}
\begin{tabular}{llrr}
\toprule
GPU & TDM & Transpose load & SDMA \\
\midrule
MI455 & Yes & Yes & Yes \\
MI355 & No & Yes & Yes \\
\bottomrule
\end{tabular}
\end{table}

\subsection{Workload selection}
For Intra-GPU data movement ISA evaluation, we focus on Compressed Sparse Attention (CSA) attention, as used in DeepSeek V4 architecture. We choose it due to 1) it is demonstrates a trend for modern LLM architecture. 2) it provides more engineering complexity, as the sparse attention mechanism requires more irregular data movement. 
The kernel configurations are shown in table \ref{tab:tdm-benchmark-matrix}.
\begin{table}[t]
\centering
\caption{Compressed Sparse Attention (CSA) config in DeepSeek V4 architecture. Every query attends to a 128-token sliding window plus the listed compressed-pool rows; T represents sequence length.}
\label{tab:tdm-benchmark-matrix}
\footnotesize
\setlength{\tabcolsep}{4pt}
\begin{tabular}{llrrl}
\toprule
Family & Ratio & Heads & Top-$k$ & Pool rows per query \\
\midrule
Flash & \texttt{cr0} & 64 & 128 & none \\
Flash & \texttt{cr4} & 64 & 640 & top-512 of $T/4$ \\
Flash & \texttt{cr128} & 64 & $128 + T/128$ & all visible of $T/128$ \\
Pro & \texttt{cr0} & 128 & 128 & none \\
Pro & \texttt{cr4} & 128 & 1152 & top-1024 of $T/4$ \\
Pro & \texttt{cr128} & 128 & $128 + T/128$ & all visible of $T/128$ \\
\bottomrule
\end{tabular}
\end{table}

Per company policy, we will only show the relative speedup. The relative speedup should be able to demonstrate the benefit of those ISA and hardware features.
\subsection{Intra-GPU data pipelining evaluation}
\subsubsection{Tensor Data Mover (TDM) feature impact analysis}
\label{sec:tdm-eval}
We evaluated our best Compressed Sparse Attention (CSA) kernel implementations with and without TDM. The non-TDM baseline loads the selected KV rows with explicit load instruction from HBM to LDS, and run two waves per SIMD to hide the load latency. (Please refer to Table \ref{tab:amd-nvidia-terminology} for the SIMD definition on AMD GPU architecture). Both versions are measured on the same MI455X with the same set of kernel configurations (flash and pro, \texttt{cr0} / \texttt{cr4} / \texttt{cr128}, $T$ = 4K / 8K / 16K). The results are shown in Table \ref{tab:tdm-impact}. The TDM-enabled design leads to on average (geometric mean) \speedupTDMAvg speedup of the training step (forward + backward) and up to \speedupTDMMax speedup; the forward alone gains 1.94$\times$ on average (1.35--2.50$\times$) and the backward 1.87$\times$ (1.34--2.90$\times$).

To explain this speedup, we further analyze the kernel difference in Table \ref{tab:tdm-kernel-compare}.
The source of speedup is the register resource TDM frees. Without TDM, the memory load exposes latency. And this latency is hidden by occupancy: two waves per SIMD, so each wave is capped at 512 VGPRs and owns 16 heads. 
With TDM, the gather writes the selected rows directly into a multi-stage LDS ring buffer and the wave only waits on \texttt{S\_WAIT\_TENSORCNT}, so the kernels drop to one wave per SIMD and give each wave up to 1024 VGPRs. This change enables a deeper pipeline and higher data reuse in the register.

MI455 architecture also introduces an asynchronous load instruction from HBM to LDS,
\texttt{GLOBAL\_LOAD\_ASYNC\_}\allowbreak\texttt{TO\_LDS}. We also tested this implementation. We were unable to achieve results better than our non-TDM baseline.

\begin{table}[t]
\centering
\caption{TDM-enabled Compressed Sparse Attention (CSA) kernel speedup over the best non-TDM implementation (higher is better).}
\label{tab:tdm-impact}
\footnotesize
\setlength{\tabcolsep}{4pt}
\begin{tabular}{llrrrr}
\toprule
Family & Ratio & $T$ & Forward & Backward & Step \\
\midrule
Flash & \texttt{cr0} & 4K & 2.34$\times$ & 2.90$\times$ & 2.74$\times$ \\
 & & 8K & 2.24$\times$ & 2.75$\times$ & 2.60$\times$ \\
 & & 16K & 2.17$\times$ & 2.75$\times$ & 2.58$\times$ \\
 & \texttt{cr4} & 4K & 1.82$\times$ & 1.56$\times$ & 1.61$\times$ \\
 & & 8K & 1.77$\times$ & 1.58$\times$ & 1.62$\times$ \\
 & & 16K & 1.76$\times$ & 1.60$\times$ & 1.63$\times$ \\
 & \texttt{cr128} & 4K & 2.50$\times$ & 2.14$\times$ & 2.22$\times$ \\
 & & 8K & 2.08$\times$ & 1.56$\times$ & 1.66$\times$ \\
 & & 16K & 1.35$\times$ & 2.19$\times$ & 1.98$\times$ \\
\midrule
Pro & \texttt{cr0} & 4K & 2.12$\times$ & 2.23$\times$ & 2.19$\times$ \\
 & & 8K & 2.06$\times$ & 2.12$\times$ & 2.11$\times$ \\
 & & 16K & 2.05$\times$ & 2.10$\times$ & 2.08$\times$ \\
 & \texttt{cr4} & 4K & 1.65$\times$ & 1.34$\times$ & 1.40$\times$ \\
 & & 8K & 1.65$\times$ & 1.35$\times$ & 1.41$\times$ \\
 & & 16K & 1.64$\times$ & 1.36$\times$ & 1.41$\times$ \\
 & \texttt{cr128} & 4K & 2.31$\times$ & 1.64$\times$ & 1.77$\times$ \\
 & & 8K & 1.88$\times$ & 1.52$\times$ & 1.60$\times$ \\
 & & 16K & 1.88$\times$ & 1.94$\times$ & 1.93$\times$ \\
\midrule
\multicolumn{3}{l}{Geometric mean} & 1.94$\times$ & 1.87$\times$ & 1.88$\times$ \\
\bottomrule
\end{tabular}
\end{table}

\begin{table}[t]
\centering
\caption{VGPRs per wave of the non-TDM and TDM CSA kernels.$^\dagger$spills VGPRs; $^\ddagger$spills on \texttt{cr128} only.}
\label{tab:tdm-kernel-compare}
\footnotesize
\setlength{\tabcolsep}{3pt}
\begin{tabular}{llrrrr}
\toprule
 & & \multicolumn{2}{c}{Forward} & \multicolumn{2}{c}{Backward} \\
\cmidrule(lr){3-4}\cmidrule(lr){5-6}
Design (waves/SIMD) & Kernel & Flash & Pro & Flash & Pro \\
\midrule
Non-TDM (2) &  & 512$^\dagger$ & 511 & 431 & 462 \\
\midrule
TDM (1) & \texttt{cr4} & 910 & 954 & 730 & 1004 \\
 & \texttt{cr0}/\texttt{cr128} & 1024 & 1024 & 1024$^\ddagger$ & 1024$^\ddagger$ \\
\bottomrule
\end{tabular}
\end{table}

\subsubsection{Transpose load ISA feature impact analysis}

\label{sec:transpose-eval}
To evaluate the transpose-load ISA, we first study the effective bandwidth of the transpose load operation. The original kernels read every transposed operand 
with \texttt{ds\_load\_tr16\_b128}. We implemented two replacements that produce the same transpose without the transpose-load ISA: One with XOR register-shuffle as shown in Table~\ref{tab:transpose-load-demo}. The other uses strided LDS loads. We measured the kernel bandwidth and show the results in Table~\ref{tab:lds-bw}. We can see that the strided LDS loads reach 0.63$\times$ of the roof-line bandwidth achieves by direct loading, and the XOR register-shuffle version achieves 0.41$\times$. The bandwidth difference is due to the micro-architecture throughput limits.

\begin{table}[t]
\centering
\caption{LDS-to-register bandwidth comparison, normalized to a plain \texttt{ds\_load\_b128} (no transpose); higher is better.}
\label{tab:lds-bw}
\footnotesize
\setlength{\tabcolsep}{4pt}
\begin{tabular}{lr}
\toprule
Load & Normalized bandwidth \\
\midrule
\texttt{ds\_load\_b128} (no transpose) & 1.00 \\
\texttt{ds\_load\_tr16\_b128} & 0.99 \\
8$\times$ strided \texttt{ds\_load\_u16} & 0.63 \\
\texttt{ds\_load\_b128} + XOR shuffle & 0.41 \\
\bottomrule
\end{tabular}
\end{table}

To evaluate the transpose-load ISA impact on the kernel level, we take the TDM-enabled kernels above and change only how transposed operands are read from LDS. 
Since the above analysis showed strided-load provides best bandwidth, we used it to replace the original transpose-load instructions \texttt{ds\_load\_tr16\_b128} in the kernels.

Table~\ref{tab:transpose-impact} shows the kernel-level result. Excluding the register spilled kernels, the transpose load ISA speeds up the training step by \speedupMatrixTransposeAvg on average (geometric mean) and by up to \speedupMatrixTransposeMax; 
The speedup is due to the transposed-load bandwidth discussed above: the transpose load provides higher bandwidth than strided-load.

The additional spills are due to the strided-load implementation: Temporary registers will be used to store the strided-load data. Those data will then be packed back. Between the loads and the packing, there will be additional registers compared to a transpose-load implementation.
This implementation is enforced by the compiler, and is the best viable implementation we can reasonably achieve. Implementing the kernel directly in assembly code might reduce the overhead, but the LDS-to-register bandwidth overhead we demonstrated in Table \ref{tab:lds-bw} is a micro-architecture limitation. Regardless, the non-spill kernel results in Table \ref{tab:transpose-impact} should demonstrate the impact of transpose-load ISA on this attention algorithm.

\begin{table}[t]
\centering
\caption{Transpose-load-ISA-enabled kernel speedup vs strided-load version (higher is better). $^\dagger$The strided-load casued spills VGPRs.}
\label{tab:transpose-impact}
\footnotesize
\setlength{\tabcolsep}{4pt}
\begin{tabular}{llrrrr}
\toprule
Family & Ratio & $T$ & Forward & Backward & Step \\
\midrule
Flash & \texttt{cr0} & 4K & 1.36$\times$ & 1.26$\times$$^\dagger$ & 1.29$\times$ \\
 & & 8K & 1.26$\times$ & 1.22$\times$$^\dagger$ & 1.23$\times$ \\
 & & 16K & 1.28$\times$ & 1.19$\times$$^\dagger$ & 1.22$\times$ \\
 & \texttt{cr4} & 4K & 1.31$\times$ & 1.16$\times$ & 1.19$\times$ \\
 & & 8K & 1.40$\times$ & 1.17$\times$ & 1.21$\times$ \\
 & & 16K & 1.41$\times$ & 1.16$\times$ & 1.21$\times$ \\
 & \texttt{cr128} & 4K & 1.39$\times$ & 3.14$\times$$^\dagger$ & 2.75$\times$ \\
 & & 8K & 1.26$\times$ & 1.12$\times$ & 1.15$\times$ \\
 & & 16K & 1.27$\times$ & 1.14$\times$ & 1.17$\times$ \\
\midrule
Pro & \texttt{cr0} & 4K & 2.15$\times$ & 1.28$\times$$^\dagger$ & 1.56$\times$ \\
 & & 8K & 2.46$\times$ & 1.27$\times$$^\dagger$ & 1.62$\times$ \\
 & & 16K & 2.49$\times$ & 1.25$\times$$^\dagger$ & 1.62$\times$ \\
 & \texttt{cr4} & 4K & 1.45$\times$ & 1.99$\times$$^\dagger$ & 1.88$\times$ \\
 & & 8K & 1.47$\times$ & 1.98$\times$$^\dagger$ & 1.88$\times$ \\
 & & 16K & 1.46$\times$ & 1.98$\times$$^\dagger$ & 1.87$\times$ \\
 & \texttt{cr128} & 4K & 1.37$\times$ & 3.30$\times$$^\dagger$ & 2.88$\times$ \\
 & & 8K & 1.29$\times$ & 2.09$\times$$^\dagger$ & 1.90$\times$ \\
 & & 16K & 1.49$\times$ & 2.05$\times$$^\dagger$ & 1.92$\times$ \\
\midrule
\bottomrule
\end{tabular}
\end{table}

\subsection{Inter-GPU data pipelining evaluation}
\label{sec:sdma_eval}
For Inter-GPU data pipelining, we will use MLperf dataset, in particular the LLama 3.1 8B, GPTOSS 20B, and FLUX.1-Schnell models as listed in Table \ref{tab:mlperf}.

\begin{table}[t]
\centering
\caption{Selected MLperf dataset.}
\label{tab:mlperf}
\begin{tabular}{llrr}
\toprule
Workload & Model & \#GPU & Scale-up/-out \\
\midrule
LLama 3.1 8B & LLM & 8 & Scale-up \\
GPTOSS 20B & LLM & 8 & Scale-up \\
FLUX.1-Schnell & Diffusion & 8 & Scale-up \\
\bottomrule
\end{tabular}
\end{table}

\subsubsection{Communication collective micro benchmark}
We first evaluate the GEMM performance when overlapped with SDMA-based or traditional Compute Unit (CU) based  RCCL communication collectives. GEMM shapes used are from Llama3 8B training. Table \ref{tab:gemm-sdma-impact} shows the relative slow down. On average, when overlapped with CU-based implementation, GEMM performance degraded by 28\% and 35\% for AllGather and ReduceScatter. But for SDMA-based implementation, the regression is only 8\% and 14\%.  This is aligned with our overhead analysis. The RCCL uses 25\% of the total Compute Units for communication. 
The higher regression for ReduceScatter is expected, as the collective relies on Compute Unit to perform the reduction. For SDMA-based ReduceScatter, the SDMA engine will move the data from remote to local, and a local persistent kernel will be used to perform the reduce operation.

\begin{table}[t]
\centering
\small
\caption{GEMM runtime increase while overlapped with collectives, relative to the same GEMM runs alone.}
\label{tab:gemm-sdma-impact}
\begin{tabular}{lrrrr}
\toprule
 & \multicolumn{2}{c}{AllGather} & \multicolumn{2}{c}{ReduceScatter} \\
\cmidrule(lr){2-3} \cmidrule(lr){4-5}
GEMM($MNK$) & CU & SDMA & CU & SDMA \\
\midrule
8k $\times$ 6k $\times$ 4k & $46\%$ & $16\%$ & $64\%$ & $15\%$ \\
8k $\times$ 4k $\times$ 4k & $29\%$ & $3.3\%$ & $44\%$ & $13\%$ \\
8k $\times$ 28k $\times$ 4k & $15\%$ & $3.6\%$ & $15\%$ & $7.3\%$ \\
8k $\times$ 4k $\times$ 14k & $23\%$ & $9.6\%$ & $23\%$ & $20\%$ \\
\midrule
Geometric mean & $28\%$ & $8.0\%$ & $35\%$ & $14\%$ \\
\bottomrule
\end{tabular}
\vspace{-2mm}
\end{table}

\subsubsection{Scale-up domain evaluation}
We compare the training throughput for LLama 3.1 8B, GPTOSS 20B and FLUX.1 models with and without SDMA based communication. The throughput speedup is shown in column $S_p$ in Table \ref{tab:sdma_workloads}. 
In the same table, we also list the communication breakdown. AG and RS columns represent the AllGather and ReduceScatter percentage of the stabilized training iteration. The sum of AG and RS represents the total percentage of overlapped communication with computation. 
Based on Amdahl's Law \cite{amdahl1967validity}, this translates into an average \speedupSDMAAvg speedup of the overlapped regions.

\begin{table}[t]
\centering
\small
\caption{Training speedup with SDMA vs. non-SDMA. AG stands for AllGather, RS for ReduceScatter, $S_p$ for Speedup. $^\dagger$ Flux workload only enabled AllGather SDMA.}
\label{tab:sdma_workloads}
\begin{tabular}{lcccc}
\toprule
Model & AG(\%)&RS(\%)&$S_p$(\%)& Regional $S_p$\\
\midrule
Llama3.1 8B & 6.2 & 9.1 & $+2.1$&1.16$\times$ \\
GPT-OSS 20B & 13.5 & 14.6 & $+4.5$& 1.18$\times$\\
FLUX.1 & {22.0$^{\dagger}$} & & $+3.1$ &1.16$\times$ \\
\bottomrule
\end{tabular}
\vspace{-1mm}
\end{table}

\subsubsection{Scale-out domain evaluation}
We are still integrating the hierarchical System and Remote DMA based communication collectives into our training workload.
To show our early results, we repeat the GEMM overlap microbenchmark on 8, 16, and 32 MI355 GPUs (i.e., one, two, and four nodes of 8 GPUs each).
The GEMM shapes are from Llama 3.1 405B training.
Table~\ref{tab:rdma_sdma_analysis} shows the relative GEMM slowdown.

With DMA-based collectives, the regression stays between 3.8\% and 4.3\%, since neither the scale-up nor the scale-out use Compute Units for data movement.

With CU-based collectives, the regression grows from 11\% on a single node to 37\% and 39\% on 16 and 32 GPUs. On 16 and 32 GPUs, this makes the overlapped GEMM on average \speedupRDMAAvg faster with hierarchical SDMA and RDMA than with CU-based collectives.

\begin{table}[t]
\centering
\small
\setlength{\tabcolsep}{3pt}
\caption{GEMM runtime increase while overlapped with AllGather on 8, 16, and 32 GPUs, relative to the same GEMM run alone.}
\label{tab:rdma_sdma_analysis}
\begin{tabular}{lrrrrrr}
\toprule
 & \multicolumn{2}{c}{8 GPUs} & \multicolumn{2}{c}{16 GPUs} & \multicolumn{2}{c}{32 GPUs} \\
\cmidrule(lr){2-3} \cmidrule(lr){4-5} \cmidrule(lr){6-7}
GEMM($MNK$) & CU & DMA & CU & DMA & CU & DMA \\
\midrule
32k$\times$16k$\times$16k & $11\%$ & $4.8\%$ & $31\%$ & $4.0\%$ & $34\%$ & $3.7\%$ \\
32k$\times$18k$\times$16k & $9.3\%$ & $3.5\%$ & $30\%$ & $3.1\%$ & $32\%$ & $3.4\%$ \\
32k$\times$16k$\times$52k & $12\%$ & $5.0\%$ & $54\%$ & $4.0\%$ & $58\%$ & $3.9\%$ \\
32k$\times$104k$\times$16k & $12\%$ & $4.0\%$ & $36\%$ & $4.0\%$ & $36\%$ & $4.2\%$ \\
\midrule
Geometric mean & $11\%$ & $4.3\%$ & $37\%$ & $3.8\%$ & $39\%$ & $3.9\%$ \\
\bottomrule
\end{tabular}
    \vspace{-2mm}
\end{table}

\section{Related works}
AMD GPU also provides other data-pipelining ISA and features, such as prefetching from HBM to different cache layers. This feature has already been utilized in our kernel implementation. TDM can move data from HBM of other GPUs in the scale-up network to the on-chip memory of the local GPU, and vise versa. Recent AMD work has enabled using TDM for tightly-coupled communication and computation, such as DeepEP and MegaMoE kernels.

\section{Future works}
Recent AMD work has been exploring reverse offloading for GPU communication, which utilizes proxy threads on the host CPU to initiate network operations. This gives us a new low overhead programming method for implementing GPU communications that are coupled with computation. We will explore this method to implement DeepEP and MegaMoE kernels. We will also continue to explore hierarchical System and Remote DMA based communication collectives and integrate it to our LLM training framework.

\section{Conclusion}
This paper presents data pipelining features of AMD GPU. The Tensor Data Mover (TDM) moves tensor data from HBM to on-chip memory. Transpose-load instructions transpose tensor data into the layout that LLM computations require as it is loaded on-chip. System DMA and Remote DMA move data between GPUs over the scale-up and scale-out networks. 

We show that these data pipelining features enables a rethinking about GPU efficiency. We need to schedule the data movement across the GPU system (e.g., from HBM to on-chip memory and between local and remote HBM) in time, so the compute units are always busy and not wasted for waiting or moving the data. It is also infeasible to prefetch everything needed to the closest memory hierarchy. The on-chip memory and HBM capacity are too small compared to the huge memory capacity needed for LLM training.

We utilize these features in our library and framework implementations for LLM training. Our evaluations show speedup on average \speedupTDMAvg for TDM, \speedupMatrixTransposeAvg for Transpose-load. For overlapped compute with communication, \speedupSDMAAvg for System DMA and \speedupRDMAAvg for Remote DMA.  

\bibliography{example_paper}
\bibliographystyle{mlsys2025}

\appendix
\section{AI tool usage}
The ROCm software, including our work, has been integrating AI tools throughout the development cycle. This includes internal and external AI models and tools. For example, Cursor, TraceLens, GEAK, Hyperloom, Claude Code. Those tools have been used to generate codes, perform testing and debugging.

For this paper, We also used AI tools to generate benchmarking scripts and schedule GPU jobs on our cluster.

For paper writing, we used the AI tool to check for grammar errors, polish language and content, and brainstorm better ways to present our ideas. 


\end{document}